\documentclass[12pt]{article}
\usepackage{amsmath}
\usepackage{amssymb}
\usepackage{graphicx}
\usepackage[numbers,sort&compress]{natbib}

\newtheorem{proposition}{Proposition}
\newtheorem{lemma}{Lemma}
\title{Dressing by regularization for the Gerdjikov-Ivanov equation and higher-order solitons}
\date{}
\author{Juanjuan Yang$^1$, Junyi Zhu$^2$\thanks{Email: jyzhu@zzu.edu.cn}~ and Linlin Wang$^2$ \\
{\small 1. School of Mathematical Sciences, Henan Institute of Science and Technology,}\\
{\small Xinxiang, Henan 453003, China}\\
{\small 2. School of Mathematics and Statistics, Zhengzhou University,}\\
{\small Zhengzhou, Henan 450001, China}}
\begin{document}
\maketitle

\begin{abstract}
Higher-order solitons, as well as simple $N$-soliton solutions, of the Gerdjikov-Ivanov equation
are derived by the dressing method based on the technique of regularization.
By the dressing transformation for the eigenfunction associated with a seed solution,
the regularity conditions of the dressed eigenfunctions are found to establish the relationship
between the potential and the scattering data.
{\bf PASC Codes:} 02.30.Ik, 02.30.Zz, 04.20.Jb
\end{abstract}

\section{Introduction}

The Gerdjikov-Ivanov (GI) equation \cite{bjp10-130}, given as
\begin{equation}\label{a1}
iq_t(x,t)+q_{xx}(x,t)-iq^2(x,t)\bar{q}_x(x,t)+\frac{1}{2}q^3(x,t)\bar{q}^2(x,t)=0,
\end{equation}
is a generalization of the derivative nonlinear Schr\"odinger equation \cite{jmp19-798,jpsj64-1519,pd25-399},
which has important applications in the fields of mathematics and physics. Here $q=q(x,t)$, $q_t=\partial q/\partial t$,
$q_{xx}=\partial^2q/\partial x^2$, and $\bar{q}$ denotes the complex conjugation of $q$.
It is also noted that the GI equation can be regarded as an extension of the nonlinear Schr\"odinger equation (NLS).
Various approaches have been proposed to construct the solitons of the GI equation such as the Hirota bilinear method \cite{jpsj64-1519},
Darboux transformation and Hamiltonian structures \cite{jmp41-7769,jmp53-033510,ps89-035501},
rogue wave \cite{jmp53-063507,ps89-035501}, algebro-geometric solutions \cite{csf22-93,jmp54-073505}, and others \cite{imrn78-4181,gmj47-99}.
The higher-order rogue wave solutions of the GI equation have been obtained by virtue of the generalized Darboux transformation \cite{ps89-035501}.

From the theoretical basis of the inverse scattering transform method, it is known that
the soliton solutions of an integrable equation are determined by the poles of the associated reflection coefficient.
Several distinct simple poles producing multisoltion solutions can be coalesced to
obtain a multiple-pole solution, if such coalescing is a regular limit \cite{ol19-619,sam110-297}.
We note that these poles are also the poles of the soliton dressing matrix,
which is a rational matrix function, and that distinct simple poles of the dressing matrix yield multiple poles
by the coalescing procedure. Thus, this limit is singular.
We reiterate that a higher-order pole in the soliton dressing matrix can not be obtained in a regular way by
coalescing simple poles in the generic mulisoliton mantrx.
In this sense, it is interesting to study the higher-order solitons.
While information regarding higher-order solitons can be found elsewhere \cite{ol19-619,sam110-297,cmp207-1,jpsj51-2029,jpsj53-2908},
these results, in contrast to those of multisoliton solutions, are scarce.
In this paper, we intend to derive the higher-order soliton solutions of the GI equation
by means of the dressing method \cite{jns20-709}. The simple $N$-soliton solutions of the GI equation are also obtained.

In the application of the dressing method, we present two cases of the dressing factor with simple poles and two-order poles, respectively.
By the dressing transformation for the eigenfunction associated with a seed solution, the regularity conditions regarding the dressed eigenfunctions are obtained,
and the relationship between the potential and the scattering data is then established. In this way, the simple $N$-soliton and the higher-order soliton solutions of
the GI equation are obtained.

\setcounter{equation}{0}
\section{Lax representation}
The GI equation (\ref{a1}) admits the following Lax representation
\begin{equation}\label{a2}
\begin{aligned}
\psi_x+ik^2\sigma_3\psi=U(x,t,k)\psi,\\
\psi_t+2ik^4\sigma_3\psi=V(x,t,k)\psi.
\end{aligned}
\end{equation}
Here $k$ is a spectral parameter, and the following definitions are applied:
\begin{equation}\label{a2a}
\begin{aligned}
U=k Q-\frac{i}{2}Q^2\sigma_3,\quad Q=\left(\begin{matrix}
0&q(x,t)\\
-\bar{q}(x,t)&0
\end{matrix}\right),\quad \sigma_3=\left(\begin{matrix}
1&0\\
0&-1
\end{matrix}\right),\\
V=2k^3Q-ik^2Q^2\sigma_3-ik Q_x\sigma_3
+\frac{1}{2}(Q_xQ-QQ_x)+\frac{i}{4}Q^4\sigma_3.
\end{aligned}\end{equation}

We assume that $q$ is a smooth potential with sufficient decay as $|x|\rightarrow\infty$,
and that the eigenfunction $\psi$ satisfies the following symmetry conditions:
\begin{equation}\label{a3}
\sigma_3\psi(x,t,-k)\sigma_3=\psi(x,t,k),\quad \psi^{-1}(x,t,k)=\psi^\dagger(x,t,\bar{k}),
\end{equation}
where the bar denotes complex conjugation. We next introduce the transformation:
\begin{equation}\label{a4}
\psi(x,t,k)=\Psi(x,t,k)e^{-i\theta\sigma_3},\quad
\theta:=k^2x+2k^4t,
\end{equation}
such that $\Psi$ satisfies the following linear system:
\begin{equation}\label{a5}
\begin{aligned}
\Psi_x+ik^2[\sigma_3,\Psi]=U(x,t,k)\Psi,\\
\Psi_t+2ik^4[\sigma_3,\Psi]=V(x,t,k)\Psi.
\end{aligned}
\end{equation}

We consider the Jost solutions $\Psi_\pm$ of the spectral equation in (\ref{a5}) obeying the asymptotic conditions
$\Psi_\pm\rightarrow I$ as $x\rightarrow\pm\infty$. Here $I$ denotes the identity matrix. Note that the first column vector denoted by $[\Psi_-]_1$ of the matrix $\Psi_-$
and the second column vector $[\Psi_+]_2$ of $\Psi_+$ are analytic in the region ${\rm Im}k^2\geq0$, and $[\Psi_+]_1$ and $[\Psi_-]_2$
are analytic in ${\rm Im}k^2\leq0$. Now, we define a matrix solution $\psi$ of (\ref{a2}) by these column vectors as
\begin{equation}\label{a6}
\psi(x,t,k)=\left\{\begin{matrix}
([\Psi_-]_1,[\Psi_+]_2)e^{-i\theta\sigma_3},&{\rm Im}k^2\geq0,\\
([\Psi_+]_1,[\Psi_-]_2)e^{-i\theta\sigma_3},&{\rm Im}k^2\leq0.
\end{matrix}\right.
\end{equation}
It is readily verifiable that the sectionally analytic function $\psi$ admits the symmetries (\ref{a3}).
We note that the eigenfunction $\psi$ defined by (\ref{a6}) is singular for $k=\infty$ and
has a jump across the contour $\mathbb{R}\cup i\mathbb{R}$.

\setcounter{equation}{0}
\section{Analysis of the eigenfunctions}
We consider a seed solution $q_0$ of the GI equation (\ref{a1}) and the corresponding eigenfunction $\psi_0$ obeying the symmetries (\ref{a3}).
In this case, the dressed eigenfunction $\psi$ can be constructed in the following form \cite{jns20-709}:
\begin{equation}\label{b1}
\psi=G\psi_0,
\end{equation}
where the dressing factor $G$ is given as
\begin{equation}\label{b2}
\begin{aligned}
G(x,t,k)=I+\sum\limits_{j=1}^N\left[\frac{A_j(x,t)}{k-k_j}-\frac{\sigma_3A_j(x,t)\sigma_3}{k+k_j}\right],\\
G^{-1}(x,t,k)=I+\sum\limits_{j=1}^N\left[\frac{A_j^\dagger(x,t)}{k-\bar{k}_j}-\frac{\sigma_3A_j^\dagger(x,t)\sigma_3}{k+\bar{k}_j}\right].
\end{aligned}
\end{equation}
We note that (\ref{b2}) is obtained by virtue of the symmetry conditions
\begin{equation}\label{b3}
\sigma_3G(x,t,-k)\sigma_3=G(x,t,k),\quad G^{-1}(x,t,k)=G^\dagger(x,t,\bar{k}),
\end{equation}
in terms of (\ref{a3}) and (\ref{b1}). One can see that $\{k_j,\bar{k}_j\}_1^N\subset\mathbb{C}$ is a collection of simple poles,
and that $A_j$ and $-\sigma_3A_j\sigma_3$ are the corresponding residues. It is apparent that $G$ and $G^{-1}$ are analytic at $k=\infty$.

Differentiation of (\ref{b1}) with respect to $x$ and $t$ implies that
\begin{equation}\label{b4}\begin{aligned}
\psi_x\psi^{-1}=G_xG^{-1}+G\psi_{0,x}\psi_0^{-1}G^{-1},\\
\psi_t\psi^{-1}=G_tG^{-1}+G\psi_{0,t}\psi_0^{-1}G^{-1}.
\end{aligned}
\end{equation}
Note that
\begin{equation}\label{b5}
\psi_{0,x}\psi^{-1}=-ik^2\sigma_3+U_0,\quad \psi_{0,t}\psi^{-1}=-2ik^4\sigma_3+V_0,\\
\end{equation}
where $U_0$ and $V_0$ are defined by (\ref{a2a}) with the seed solution $q_0$ or $Q_0$.
It is worthwhile to mention that $N$-solitonic solution of the focusing nonlinear Schr\"odinger equation with nonvanishing boundary conditions
has been derived in Ref.\cite{non27-r1}.

It follows from equations (\ref{b4}) and (\ref{b5}) that $\psi_x\psi^{-1}$ and $\psi_t\psi^{-1}$ are singular at the points
$k=\infty$ and $\{k_j,\bar{k}_j\}_1^N$.
To investigate the singularity at $k=\infty$, we consider the asymptotic behavior of $G$ and $G^{-1}$ as $k\rightarrow\infty$, and let
\begin{equation}\label{b6}
\begin{aligned}
G=I+k^{-1}G_1+k^{-2}G_2+k^{-3}G_3+O(k^{-4}),\\
G^{-1}=I+k^{-1}G^{(1)}+k^{-2}G^{(2)}+k^{-3}G^{(3)}+O(k^{-4}),
\end{aligned}
\end{equation}
where $O(k^{-4})$ is of order $k^{-4}$ as $k\rightarrow\infty$, and
\begin{equation}\label{b7}
\begin{aligned}
G_1=\sum\limits_{j=1}^N(A_j-\sigma_3A_j\sigma_3),\\
G_2=\sum\limits_{j=1}^Nk_j(A_j+\sigma_3A_j\sigma_3),\\
G_3=\sum\limits_{j=1}^Nk_j^2(A_j-\sigma_3A_j\sigma_3),\\
\cdots\quad\cdots\quad\cdots
\end{aligned}
\end{equation}
and
\begin{equation}\label{b8}
\begin{aligned}
G^{(1)}=-G_1,\quad G^{(2)}=-G_2-G_1G^{(1)},\\
G^{(3)}=-G_3-G_2G^{(1)}-G_1G^{(2)}.
\end{aligned}
\end{equation}
It is noted from (\ref{b7}) that the matrices $G_{2l-1}$ and $G_{2l},(l=1,2,\cdots)$ are off-diagonal and diagonal, respectively.

Substitution of (\ref{b6}) and (\ref{b5}) into (\ref{b4}) implies the following asymptotic behaviors as $k\rightarrow\infty$:
\begin{equation}\label{b9}
\begin{aligned}
\psi_x\psi^{-1}&=-ik^2\sigma_3+Qk-\frac{i}{2}Q^2\sigma_3+O(k^{-1}),\quad Q=Q_0+2i\sigma_3G_{1},\\
\psi_t\psi^{-1}&=-2ik^4\sigma_3+2k^3Q-ik^2Q^2\sigma_3+Q^{(1)}k+Q^{(0)}+O(k^{-1}),
\end{aligned}
\end{equation}
in view of (\ref{b8}), where $Q^{(1)}$ and $Q^{(0)}$ are some certain functions to be determined. Hence, the representations
\begin{equation}\label{b10}
\begin{aligned}
\psi_x\psi^{-1}&+ik^2\sigma_3-Qk+\frac{i}{2}Q^2\sigma_3,\\
\psi_t\psi^{-1}&+2ik^4\sigma_3-2k^3Q+ik^2Q^2\sigma_3-Q^{(1)}k-Q^{(0)},
\end{aligned}
\end{equation}
are analytic near $k=\infty$. In other words, these representations are analytic on the entire Riemann $k$-sphere except possibly at
points in the set $\{\pm k_j,\pm\bar{k}_j\}_1^N$. However, these singularities can be eliminated by the proper selection of some $A_j$ in (\ref{b7})
and $Q^{(1)}, Q^{(0)}$.
We now assume that such a selection has been made, which means that the functions in (\ref{b10}) are analytic on the entire Riemann $k$-sphere.
In this case, because the representations in (\ref{b10}) tend to zero as $k\rightarrow\infty$, it follows from Liouville's theorem that they both vanish identically.
Thus, we have the following pair of equations:
\begin{equation}\label{b11}
\begin{aligned}
\psi_x\psi^{-1}&+ik^2\sigma_3-Qk+\frac{i}{2}Q^2\sigma_3=0,\\
\psi_t\psi^{-1}&+2ik^4\sigma_3-2k^3Q+ik^2Q^2\sigma_3-Q^{(1)}k-Q^{(0)}=0.
\end{aligned}
\end{equation}
It must be mentioned that, while the functions $Q^{(1)}$ and $Q^{(0)}$ are properly selected, their identity remains unknown.
To identify these functions, we retain the assumption that $A_j$ in (\ref{b7}) are chosen properly.
In this case, the expansions in (\ref{b9}) remain valid, but the term $O(k^{-1})$ is zero, which implies that $Q^{(1)}=-iQ_x\sigma_3$ and
$Q^{(0)}=(Q_xQ-QQ_x)/2+i(Q^4/4)\sigma_3$. Hence, we have the Lax representation (\ref{a2}).
In addition, from (\ref{b9}) and (\ref{b7}), we obtain the solution of the GI equation:
\begin{equation}\label{b12}
q=q_0+4i\sum\limits_{j=1}^N(A_j)_{12}.
\end{equation}

\setcounter{equation}{0}
\section{The dressing transformation}
In this section, we determine $A_j$ in (\ref{b7}) ensuring that the functions $\psi_x\psi^{-1}$
and $\psi_t\psi^{-1}$ are regular at the points $\{\pm k_j,\pm\bar{k}_j\}_1^N$.
To this end, we consider the following series of $N$ consecutive dressing transformations, each of which adds two poles \cite{jns20-709}:
\begin{equation}\label{b13}
G=D_ND_{N-1}\cdots D_1,\quad \psi_j=D_j\psi_{j-1},\quad j=1,2,\cdots,N,
\end{equation}
where the eigenfunction $\psi_0$ is chosen to be regular and
\begin{equation}\label{b14}
\begin{aligned}
D_j=I+\frac{B_j(x,t)}{k-k_j}-\frac{\sigma_3B_j(x,t)\sigma_3}{k+k_j},\\
D_j^{-1}=I+\frac{B_j^\dagger(x,t)}{k-\bar{k}_j}-\frac{\sigma_3B_j^\dagger(x,t)\sigma_3}{k+\bar{k}_j}.
\end{aligned}
\end{equation}

The regularization of the functions $\psi_{j,x}\psi_j^{-1}$
and $\psi_{j,t}\psi_j^{-1}$ for $j=1,2,\cdots,N$ can proceed by induction. Supposing that the functions $\psi_{j-1,x}\psi_{j-1}^{-1}$,
and $\psi_{j-1,t}\psi_{j-1}^{-1}$ are analytic at the points $\{\pm k_l, \pm\bar{k}_l\}_1^{j-1}$,
we show that $\psi_{j,x}\psi_{j}^{-1}$ and $\psi_{j,t}\psi_{j}^{-1}$
are analytic at the points $\{\pm k_l, \pm\bar{k}_l\}_1^{j}$, where $\psi_j=D_j\psi_{j-1}$.
To this end, we must differentiate the equation $\psi_j=D_j\psi_{j-1}$ with respect to $x$, which provides
\begin{equation}\label{b15}
\psi_{j,x}\psi_j^{-1}=D_{j,x}D_j^{-1}+D_j\psi_{j-1,x}\psi_{j-1}^{-1}D_j^{-1}.
\end{equation}
The assumption regarding $\psi_{j-1}$, together with (\ref{b14}), implies that the right-hand side of (\ref{b15}) is analytic
at the points $\{\pm k_l, \pm\bar{k}_l\}_1^{j-1}$. We must therefore determine the conditions ensuring that $\psi_{j,x}\psi_j^{-1}$ is regular at the simple poles
$\pm k_j$ and $\pm\bar{k}_j$. We note that the residue of  $\psi_{j,x}\psi_j^{-1}$ at the point $k_j$ is given by
\begin{equation}\label{b16}
\begin{aligned}
{\rm Res}[\psi_{j,x}\psi_j^{-1}, k_j]=\left(B_j\psi_{j-1}(k_j)\right)_x\psi_{j-1}^{-1}(k_j)D_j^{-1}(k_j).
\end{aligned}
\end{equation}
For algorithmic convenience, we take
\begin{equation}\label{b17}
B_j(x,t)=|z_j(x,t)\rangle\langle y_j(x,t)|,
\end{equation}
where $|z_j\rangle=\langle z_j|^\dagger$ is a column vector and $\langle y_j|=|y_j\rangle^\dagger$ a row vector. Hence, equation
(\ref{b16}) reduces to
\begin{equation}\label{b18}
\begin{aligned}
{\rm Res}[\psi_{j,x}\psi_j^{-1}, k_j]=&|z_j\rangle_x\langle y_j|D_j^{-1}(k_j)\\
&+|z_j\rangle\left(\langle y_j|\psi_{j-1}(k_j)\right)_x\psi_{j-1}^{-1}(k_j)D_j^{-1}(k_j).
\end{aligned}
\end{equation}
Now, if we take
\begin{equation}\label{b19}
\langle y_j|D_j^{-1}(k_j)=0,
\end{equation}
and
\begin{equation}\label{b20}
\left(\langle y_j|\psi_{j-1}(k_j)\right)_x=0,
\end{equation}
 then $\psi_{j,x}\psi_j^{-1}$ is regular at the point  $k_j$.
Similar regularization of $\psi_{j,t}\psi_j^{-1}$ at the point $k_j$ requires conditions (\ref{b19}) and
$\left(\langle y_j|\psi_{j-1}(k_j)\right)_t=0$. The latter condition, together with (\ref{b20}), implies that
$\langle y_j(x,t)|$ can be defined as
\begin{equation}\label{b21}
\langle y_j|=\beta_j\psi_{j-1}^{-1}(k_j),
\end{equation}
where $\beta_j$ is an arbitrary constant row vector. Thus, to construct the concrete expression of $B_j$ in (\ref{b17}),
one must determine the representation of $|z_j(x,t)\rangle$.
We note that condition (\ref{b19}), together with (\ref{b14}) and (\ref{b17}), produces
$$\begin{aligned}
\langle y_j|+\frac{\langle y_j|y_j\rangle}{k_j-\bar{k}_j}\langle z_j|
-\frac{\langle y_j|\sigma_3|y_j\rangle}{k_j+\bar{k}_j}\langle z_j|\sigma_3=0.
\end{aligned}$$
The Hermitian conjugation of this equation implies that
\begin{equation}\label{b22}
\begin{aligned}
(I+\sigma_3)|y_j\rangle=\frac{2}{k_j^2-\bar{k}_j^2}\alpha_j^{-1}(I+\sigma_3)|z_j\rangle,\\
(I-\sigma_3)|y_j\rangle=\frac{2}{k_j^2-\bar{k}_j^2}\bar\alpha_j^{-1}(I-\sigma_3)|z_j\rangle,
\end{aligned}\end{equation}
where $\alpha_j$ is defined by
\begin{equation}\label{b23}
\alpha_j^{-1}=\langle y_j|Z|y_j\rangle, \quad Z={\rm diag}(k_j,\bar{k}_j).
\end{equation}
It follows from (\ref{b22}) that $|z_j(x,t)\rangle$ takes the form
\begin{equation}\label{b24}
|z_j\rangle=\frac{k_j^2-\bar{k}_j^2}{2}\Lambda|y_j\rangle, \quad \Lambda={\rm diag}(\alpha_j,\bar\alpha_j).
\end{equation}
We note that the $2\times2$ matrix-valued function $B_j(x,t)$ is defined by (\ref{b17}), (\ref{b21}) and (\ref{b24}).

With $B_j(x,t)$ in hand, one finds, from the second equation of (\ref{b14}), that
\begin{equation}\label{b25}
\begin{aligned}
D_j^{-1}(k_j)-\sigma_3D_j^{-1}(k_j)\sigma_3=\frac{2k_j}{k_j^2-\bar{k}_j^2}\left(B_j^\dagger-\sigma_3B_j^\dagger\sigma_3\right),\\
D_j^{-1}(k_j)+\sigma_3D_j^{-1}(k_j)\sigma_3=2+\frac{2k_j}{k_j^2-\bar{k}_j^2}\left(B_j^\dagger+\sigma_3B_j^\dagger\sigma_3\right),\\
\end{aligned}
\end{equation}
which further imply that
\begin{equation}\label{b26}
\sigma_2B_j^T(x,t)\sigma_2=\frac{k_j^2-\bar{k}_j^2}{2k_j}D_j^{-1}(x,t,k_j),\quad \sigma_2=\left(\begin{matrix}
0&-i\\
i&0
\end{matrix}\right),
\end{equation}
in terms of the definition (\ref{b23}). Moreover, it follows from the first equation of (\ref{b14}) that
\begin{equation}\label{b27}
\begin{aligned}
&\sigma_2D_j^T(x,t,k)\sigma_2=\frac{k^2-\bar{k}_j^2}{k^2-k_j^2}D_j^{-1}(x,t,k).
\end{aligned}
\end{equation}

We now seek to determine $A_j$ in (\ref{b7}). We note that $A_j$ is the residue of $G(k)$ in (\ref{b2}) at $k_j$.
Thus, from (\ref{b13}), one obtains
\begin{equation}\label{b28}
A_j=D_N(k_j)\cdots D_{j+1}(k_j)B_jD_{j-1}(k_j)\cdots D_1(k_j).
\end{equation}
Equation (\ref{b28}), together with (\ref{b26}) and (\ref{b27}), implies that
\begin{equation}\label{b29}
A_j=a_j^{-1}\sigma_2G^{-1}(x,t,k_j)^T\sigma_2, \quad a_j=\frac{2k_j}{k_j^2-\bar{k}_j^2}
\prod\limits_{k\neq j}\frac{k_j^2-k_k^2}{k_j^2-\bar{k}_k^2}.
\end{equation}
On the other hand, we note that the definition $\psi_j(k)=D_j(k)\psi_{j-1}(k)$ in (\ref{b13})
implies $\psi_{j-1}^{-1}(k)=\psi_j^{-1}(k)D_j(k)$. Hence, equation (\ref{b28}) can be written as
\begin{equation}\label{b30}
A_j=D_N(k_j)\cdots D_{j+1}(k_j)B_j\psi_{j-1}(k_j)\psi_0^{-1}(k_j),
\end{equation}
which can be further rewritten as
\begin{equation}\label{b31}
A_j=\frac{1}{a_j}\left(\begin{aligned}
r_j(x,t)\\
s_j(x,t)
\end{aligned}\right)(b_j,b_j^{-1})\psi_0^{-1}(k_j),
\end{equation}
in terms of (\ref{b17}). Here, the constant row vector $\beta_j$ in (\ref{b21}) is chosen as $(b_j,b_j^{-1})$,
and the functions $r_j(x,t)$ and $s_j(x,t)$ can be determined by the following system:
\begin{equation}\label{b32}
\begin{aligned}
\sigma_2G^{-1}(x,t,k_j)^T\sigma_2=\left(\begin{aligned}
r_j(x,t)\\
s_j(x,t)
\end{aligned}\right)(b_j,b_j^{-1})\psi_0^{-1}(x,t,k_j),\\
G(x,t,\bar{k}_l)=\sigma_2\left(\begin{aligned}
\overline{r_l(x,t)}\\
\overline{s_l(x,t)}
\end{aligned}\right)(\bar{b}_l,\bar{b}_l^{-1})\overline{\psi_0^{-1}(k_l)}\sigma_2,
\end{aligned}
\end{equation}
and
\begin{equation}\label{b33}
G(x,t,\bar{k}_l)=I+\sum\limits_{j=1}^N\frac{1}{a_j}\left(
\frac{\sigma_2G^{-1}(x,t,k_j)^T\sigma_2}{\bar{k}_l-k_j}
-\frac{\sigma_3\sigma_2G^{-1}(x,t,k_j)^T\sigma_2\sigma_3}{\bar{k}_l+k_j}\right).
\end{equation}
\begin{proposition}\label{p2}
Let $N$ be a positive integer and let $\{k_j,b_j\}_1^N$ be nonzero complex constants such that
$k_j\neq k_l$ for $j\neq l$. Assume that $q_0$ satisfies (\ref{a1}), and let $\psi_0(x,t,k)$
be an associated eigenfunction obeying the symmetries (\ref{b3}). Then, the following function $q(x,t)$
is also a solution of (\ref{a1}):
\begin{equation}\label{b34}
q(x,t)=q_0+4i\sum\limits_{j=1}^N(A_j)_{12},
\end{equation}
where $A_j$ is given by (\ref{b31}), with the functions $\{r_j(x,t),s_j(x,t)\}_1^N$ determined by
the linear algebraic system $(l=1,\cdots,N)$
\begin{equation}\label{b35}
\begin{aligned}
&\sigma_2\overline{\psi_0(k_l)}\left(\begin{aligned}
\bar{b}_l^{-1}\\
-\bar{b}_l
\end{aligned}\right)+\sum\limits_{j=1}^N\frac{1}{a_j}\left[\frac{1}{\bar{k}_l-k_j}\left(\begin{aligned}
r_j\\
s_j
\end{aligned}\right)(b_j,b_j^{-1})\psi_0^{-1}\right.\\
&\qquad\left.-\frac{1}{\bar{k}_l+k_j}\sigma_3\left(\begin{aligned}
r_j\\
s_j
\end{aligned}\right)(b_j,b_j^{-1})\psi_0^{-1}\sigma_3\right]\sigma_2\bar{k}_l\left(\begin{aligned}
\bar{b}_l^{-1}\\
-\bar{b}_l
\end{aligned}\right)=0.
\end{aligned}
\end{equation}
\end{proposition}

In particular, we let the seed solution $q_0=0$ and the corresponding eigenfunction $\psi_0=\exp(-i\theta\sigma_3)$,
where $\theta$ is defined in (\ref{a4}).
It then follows from {\bf Proposition} \ref{p2} that the $N$-soliton solution of the GI equation (\ref{a1}) takes the form
\begin{equation}\label{b25}
q=2i\sum\limits_{l,j=1}^N\bar{p_l}^2\left(K^{-1}\right)_{lj},\quad p_l=b_le^{i(k_l^2x+k_l^4t)},
\end{equation}
where the entries of the $N\times N$ matrix $K=K(x,t)$ are defined by
$$K_{jl}=\frac{1}{k_j^2-\bar{k}_l^2}(k_jp_j^2\bar{p_l}^2+\bar{k}_l).$$

\setcounter{equation}{0}
\section{Analysis of the eigenfunctions with multiple poles}
In this section, we extend the approach used in the above two sections to study higher-order solitons of the GI equation.
For convenience, we consider only the case of two-order poles.
Let $q_0$ be a seed solution  of the GI equation (\ref{a1}) and $\psi_0$ be the corresponding eigenfunction obeying the symmetries (\ref{b3}),
and then seek a dressed eigenfunction $\psi$ of the form
\begin{equation}\label{c1}
\psi=\tilde{G}\psi_0.
\end{equation}
Here the matrix $\tilde{G}$ has the form
\begin{equation}\label{c2}
\begin{aligned}
\tilde{G}(x,t,k)=&I+\sum\limits_{j=1}^N\left[\frac{A_j(x,t)}{k-k_j}-\frac{\sigma_3A_j(x,t)\sigma_3}{k+k_j}\right]\\
&\quad+\sum\limits_{j=1}^N\left[\frac{\tilde{A}_j(x,t)}{(k-k_j)^2}+\frac{\sigma_3\tilde{A}_j(x,t)\sigma_3}{(k+k_j)^2}\right],\\
\tilde{G}^{-1}(x,t,k)=&I+\sum\limits_{j=1}^N\left[\frac{A_j^\dagger(x,t)}{k-\bar{k}_j}-\frac{\sigma_3A_j^\dagger(x,t)\sigma_3}{k+\bar{k}_j}\right]\\
&\quad+\sum\limits_{j=1}^N\left[\frac{\tilde{A}_j^\dagger(x,t)}{(k-\bar{k}_j)^2}+\frac{\sigma_3\tilde{A}_j^\dagger(x,t)\sigma_3}{(k+\bar{k}_j)^2}\right],
\end{aligned}
\end{equation}
in terms of the symmetry conditions
\begin{equation}\label{c3}
\sigma_3\tilde{G}(x,t,-k)\sigma_3=\tilde{G}(x,t,k),\quad \tilde{G}^{-1}(x,t,k)=\tilde{G}^\dagger(x,t,\bar{k}),
\end{equation}
obtained from (\ref{b3}) and (\ref{c1}). We note that $\{k_j,\bar{k}_j\}_1^N\subset\mathbb{C}$ is a collection of poles of order 2.
Then, the functions $\psi_x\psi^{-1}$ and $\psi_t\psi^{-1}$ can be derived similarly as in (\ref{b4}), which have the same singularity
at $\infty$, but have two-order singularities at the points $\{\pm k_j,\pm\bar{k}_j\}_1^N$.

According to a similar discussion concerning the regularization of the functions $\psi_x\psi^{-1}$
and $\psi_t\psi^{-1}$ at $k=\infty$ conducted in Section 2, we find, under a proper assumption regarding $A_j$ and $\tilde{A}_j$, that
\begin{equation}\label{c4}
 Q=Q_0+2i\sigma_3\tilde{G}_{1},
\end{equation}
where $\tilde{G}_{j}, (j=1,2,\cdots)$ are defined as in (\ref{b6}), but, here, they are specifically
\begin{equation}\label{c5}
\begin{aligned}
\tilde{G}_{1}&=\sum\limits_{j=1}^N(A_j-\sigma_3A_j\sigma_3),\\
\tilde{G}_{2}&=\sum\limits_{j=1}^Nk_j(A_j+\sigma_3A_j\sigma_3)+\sum\limits_{j=1}^N(\tilde{A}_j+\sigma_3\tilde{A}_j\sigma_3),\\
\tilde{G}_{3}&=\sum\limits_{j=1}^Nk_j^2(A_j-\sigma_3A_j\sigma_3)+2\sum\limits_{j=1}^Nk_j(\tilde{A}_j-\sigma_3\tilde{A}_j\sigma_3),\\
&\cdots\quad\cdots\quad\cdots
\end{aligned}
\end{equation}
To finish the regularization, we find  $A_j$ and $\tilde{A}_j$ in (\ref{c5}) to ensure that the functions $\psi_{j,x}\psi_j^{-1}$
and $\psi_{j,t}\psi_j^{-1}$ are regular at the points $\pm k_j$ and $\pm\bar{k}_j$.
To this end, we consider the following series of $N$ consecutive dressing transformations,
each of which adds two poles of order 2:
\begin{equation}\label{d1}
G=D_ND_{N-1}\cdots D_1,\quad \psi_j=D_j\psi_{j-1}.
\end{equation}
Here,
\begin{equation}\label{d2}
D_j(k)=I+\frac{B_j(x,t)}{k-k_j}-\frac{\sigma_3B_j(x,t)\sigma_3}{k+k_j}+\frac{\tilde{B}_j(x,t)}{(k-k_j)^2}
+\frac{\sigma_3\tilde{B}_j(x,t)\sigma_3}{(k+k_j)^2},
\end{equation}
\begin{equation}\label{d3}
D_j^{-1}(k)=I+\frac{B_j^\dagger(x,t)}{k-\bar{k}_j}-\frac{\sigma_3B_j^\dagger(x,t)\sigma_3}{k+\bar{k}_j}
+\frac{\tilde{B}_j^\dagger(x,t)}{(k-\bar{k}_j)^2}+\frac{\sigma_3\tilde{B}_j^\dagger(x,t)\sigma_3}{(k+\bar{k}_j)^2},
\end{equation}
where $B_j(x,t)$ and $\tilde{B}_j(x,t)$ are $2\times2$ matrix-valued functions.

\begin{proposition}\label{p3}
Let $\Upsilon_j$ and $\tilde\Upsilon_j$ be nonzero constant row vectors. Define the row vectors $\langle y_j|,\langle\tilde{y}_j|$ in terms of the $(j-1)th$
eigenfunctions $\psi_{j-1}$ by
\begin{equation}\label{d4}
\langle y_j|=\Upsilon_j\psi_{j-1}^{-1}(k_j),\quad
\langle\tilde{y}_j|=\tilde\Upsilon_j\psi_{j-1}^{-1}(k_j)+\Upsilon_j(\psi_{j-1}^{-1})^\prime(k_j),
\end{equation}
where
\begin{equation}\label{d5}
(\psi_{j-1}^{-1})^\prime(k_j)=\frac{{\rm d}}{{\rm d}k}\psi^{-1}_{j-1}(k)|_{k=k_j}.
\end{equation}
Given these vectors and the points $k_j$ and $\bar{k}_j$, we define a set of scalar functions:
\begin{equation}\label{d6}
\begin{aligned}
a_{j,11}^\pm=k_j\langle y_j|(I\pm\sigma_3)|y_j\rangle +\bar{k}_j\langle y_j|(I\mp\sigma_3)|y_j\rangle,\\
a_{j,12}^\pm=k_j\langle y_j|(I\pm\sigma_3)|\tilde{y}_j\rangle +\bar{k}_j\langle y_j|(I\mp\sigma_3)|\tilde{y}_j\rangle,\\
a_{j,21}^\pm=k_j\langle \tilde{y}_j|(I\pm\sigma_3)|y_j\rangle +\bar{k}_j\langle \tilde{y}_j|(I\mp\sigma_3)|y_j\rangle,\\
a_{j,22}^\pm=k_j\langle \tilde{y}_j|(I\pm\sigma_3)|\tilde{y}_j\rangle +\bar{k}_j\langle \tilde{y}_j|(I\mp\sigma_3)|\tilde{y}_j\rangle,\\
b_{j,11}^\pm=(k_j^2+\bar{k}_j^2)\langle y_j|(I\pm\sigma_3)|y_j\rangle +2k_j\bar{k}_j\langle y_j|(I\mp\sigma_3)|y_j\rangle,\\
b_{j,12}^\pm=(k_j^2+\bar{k}_j^2)\langle y_j|(I\pm\sigma_3)|\tilde{y}_j\rangle +2k_j\bar{k}_j\langle y_j|(I\mp\sigma_3)|\tilde{y}_j\rangle,\\
b_{j,21}^\pm=(k_j^2+\bar{k}_j^2)\langle \tilde{y}_j|(I\pm\sigma_3)|y_j\rangle +2k_j\bar{k}_j\langle \tilde{y}_j|(I\mp\sigma_3)|y_j\rangle,\\
c_{j,11}^\pm=(k_j^3+3k_j\bar{k}_j^2)\langle y_j|(I\pm\sigma_3)|y_j\rangle +(\bar{k}_j^3+3\bar{k}_jk_j^2)\langle y_j|(I\mp\sigma_3)|y_j\rangle,\\
\end{aligned}
\end{equation}
where $|y_j\rangle=\langle y_j|^\dagger, |\tilde{y}_j\rangle=\langle \tilde{y}_j|^\dagger$.
Define the functions $B_j(x,t)$ and $\tilde{B}_j(x,t)$ as
\begin{equation}\label{d7}
B_j=(|z_j\rangle,|\tilde{z}_j\rangle)\left(\begin{aligned}
\langle\tilde{y}_j|\\
\langle y_j|
\end{aligned}\right),\quad
\tilde{B}_j=(|z_j\rangle,|\tilde{z}_j\rangle)\sigma_+\left(\begin{aligned}
\langle\tilde{y}_j|\\
\langle y_j|
\end{aligned}\right),\quad \sigma_+=\left(\begin{matrix}
0&1\\
0&0
\end{matrix}\right),
\end{equation}
where the column vectors $|z_j\rangle$ and $|\tilde{z}_j\rangle$  are define as
\begin{equation}\label{d8}
|z_j\rangle=\Lambda_j|\tilde{y}_j\rangle+\Omega_j|y_j\rangle,\quad
|\tilde{z}_j\rangle=\tilde\Lambda_j|\tilde{y}_j\rangle+\tilde\Omega_j|y_j\rangle.
\end{equation}
Here the $2\times2$ matrices $\Lambda_j,\tilde\Lambda_j,\Omega_j$ and $\tilde\Omega_j$ are diagonal ones, denoted conveniently as ${\rm diag}(\circledast_1,\circledast_2)$.
The entries of these diagonal matrices are defined as
\begin{equation}\label{d9}
\begin{aligned}
\Lambda_{j,1}=-\frac{(k_j^2-\bar{k}_j^2)^3}{\Delta_j^+}a_{j,11}^+,\quad \Lambda_{j,2}=-\frac{(k_j^2-\bar{k}_j^2)^3}{\Delta_j^-}a_{j,11}^-,\\
\Omega_{j,1}=\frac{(k_j^2-\bar{k}_j^2)^2}{\Delta_j^+}[(k_j^2-\bar{k}_j^2)a_{j,12}^++b_{j,11}^-],\quad
\Omega_{j,2}=\frac{(k_j^2-\bar{k}_j^2)^2}{\Delta_j^-}[(k_j^2-\bar{k}_j^2)a_{j,12}^-+b_{j,11}^+],\\
\tilde\Lambda_{j,1}=\frac{(k_j^2-\bar{k}_j^2)^2}{\Delta_j^+}[(k_j^2-\bar{k}_j^2)a_{j,21}^+-b_{j,11}^+],\quad
\tilde\Lambda_{j,2}=\frac{(k_j^2-\bar{k}_j^2)^2}{\Delta_j^-}[(k_j^2-\bar{k}_j^2)a_{j,21}^--b_{j,11}^-],\\
\tilde\Omega_{j,1}=\frac{(k_j^2-\bar{k}_j^2)}{\Delta_j^+}[-(k_j^2-\bar{k}_j^2)^2a_{j,22}^++(k_j^2-\bar{k}_j^2)(b_{j,12}^+-b_{j,21}^-)+2c_{j,11}^-],\\
\tilde\Omega_{j,2}=\frac{(k_j^2-\bar{k}_j^2)}{\Delta_j^-}[-(k_j^2-\bar{k}_j^2)^2a_{j,22}^-+(k_j^2-\bar{k}_j^2)(b_{j,12}^--b_{j,21}^+)+2c_{j,11}^+],
\end{aligned}
\end{equation}
where
\begin{equation}\label{d10}
\begin{aligned}
\Delta_j^\pm=&(k_j^2-\bar{k}_j^2)^2(a_{j,21}^\pm a_{j,12}^\pm-a_{j,11}^\pm a_{j,22}^\pm)+2a_{j,11}^\pm c_{j,11}^\mp-b_{j,11}^\pm b_{j,11}^\mp\\
&+(k_j^2-\bar{k}_j^2)[a_{j,21}^\pm b_{j,11}^\mp-a_{j,12}^\pm b_{j,11}^\pm+a_{j,11}^\pm(b_{j,12}^\pm-b_{j,21}^\mp)].
\end{aligned}
\end{equation}
Define $\psi=\psi_N$ according to (\ref{d1})$-$(\ref{d10}). Then, the functions $\psi_x\psi^{-1}$ and $\psi_t\psi^{-1}$ are
analytic at the points in the set $\{\pm k_j, \pm\bar{k}_j\}_1^N$.
\end{proposition}
Proof: This proposition can be also proven by induction. Supposing that $\psi_{j-1}$ has been defined according to (\ref{d1})$-$(\ref{d10}) and that the functions
$\psi_{j-1,x}\psi_{j-1}^{-1}$ and $\psi_{j-1,t}\psi_{j-1}^{-1}$ are analytic at the set of points
$\{\pm k_l, \pm\bar{k}_l\}_1^{j-1}$, we shall show that $\psi_{j,x}\psi_{j}^{-1}$ and $\psi_{j,t}\psi_{j}^{-1}$
are analytic at the points $\{\pm k_l, \pm\bar{k}_l\}_1^{j}$, where $\psi_j=D_j\psi_{j-1}$.
To this end, we must differentiate the equation $\psi_j=D_j\psi_{j-1}$ with respect to $x$, which provides
\begin{equation}\label{d11}
\psi_{j,x}\psi_j^{-1}=D_{j,x}D_j^{-1}+D_j\psi_{j-1,x}\psi_{j-1}^{-1}D_j^{-1}.
\end{equation}
The assumption on $\psi_{j-1}$ together with (\ref{d2}) and (\ref{d3}) implies that the right-hand side of (\ref{d11}) is analytic
at the points $\{\pm k_l, \pm\bar{k}_l\}_1^{j-1}$. We must therefore show that $\psi_{j,x}\psi_j^{-1}$ is regular at points
$\pm k_j$ and $\pm\bar{k}_j$. Firstly, we note that $\psi_{j,x}\psi_j^{-1}$ has the following asymptotic behavior near the point
$k_j$:
\begin{equation}\label{d12}
\begin{aligned}
\psi_{j,x}\psi_j^{-1}\sim&\frac{1}{(k-k_j)^2}\left\{(|z_j\rangle,|\tilde{z}_j\rangle)_x\sigma_+\left(\begin{aligned}
\langle\tilde{y}_j|\\
\langle y_j|
\end{aligned}\right)D_j^{-1}(k_j)\right.\\
&\quad\left.+(|z_j\rangle,|\tilde{z}_j\rangle)\sigma_+\left[\left(\begin{aligned}
\langle\tilde{y}_j|\\
\langle y_j|
\end{aligned}\right)\psi_{j-1}(k_j)\right]_x\psi_{j-1}^{-1}(k_j)D_j^{-1}(k_j)\right\}\\
&+\frac{1}{k-k_j}\left\{(|z_j\rangle,|\tilde{z}_j\rangle)_x\left[\sigma_+\left(\begin{aligned}
\langle\tilde{y}_j|\\
\langle y_j|
\end{aligned}\right)(D_j^{-1})^\prime(k_j)+\left(\begin{aligned}
\langle\tilde{y}_j|\\
\langle y_j|
\end{aligned}\right)D_j^{-1}(k_j)\right]\right.\\
&+(|z_j\rangle,|\tilde{z}_j\rangle)\left[\left(\begin{aligned}
\langle\tilde{y}_j|\\
\langle y_j|
\end{aligned}\right)\psi_{j-1}(k_j)+\sigma_+\left(\begin{aligned}
\langle\tilde{y}_j|\\
\langle y_j|
\end{aligned}\right)(\psi_{j-1})^\prime(k_j)\right]_x\psi_{j-1}^{-1}(k_j)D_j^{-1}(k_j)\\
&\left.+(|z_j\rangle,|\tilde{z}_j\rangle)\sigma_+\left[\left(\begin{aligned}
\langle\tilde{y}_j|\\
\langle y_j|
\end{aligned}\right)\psi_{j-1}(k_j)\right]_x\frac{{\rm d}}{{\rm d}k}[\psi_{j-1}^{-1}(k)D_j^{-1}(k)]_{k=k_j}\right\}+O(1),
\end{aligned}
\end{equation}
where the definition of $(D_j^{-1})^\prime(k_j)$ and $(\psi_{j-1})^\prime(k_j)$ are similar to that given in (\ref{d5}). If we let
\begin{equation}\label{d13}\begin{aligned}
\sigma_+\left[\left(\begin{aligned}
\langle\tilde{y}_j|\\
\langle y_j|
\end{aligned}\right)\psi_{j-1}(k_j)\right]_x=0,\\
\left[\left(\begin{aligned}
\langle\tilde{y}_j|\\
\langle y_j|
\end{aligned}\right)\psi_{j-1}(k_j)+\sigma_+\left(\begin{aligned}
\langle\tilde{y}_j|\\
\langle y_j|
\end{aligned}\right)(\psi_{j-1})^\prime(k_j)\right]_x=0,
\end{aligned}
\end{equation}
and
\begin{equation}\label{d14}\begin{aligned}
\sigma_+\left(\begin{aligned}
\langle\tilde{y}_j|\\
\langle y_j|
\end{aligned}\right)D_j^{-1}(k_j)=0,\\
\sigma_+\left(\begin{aligned}
\langle\tilde{y}_j|\\
\langle y_j|
\end{aligned}\right)(D_j^{-1})^\prime(k_j)+\left(\begin{aligned}
\langle\tilde{y}_j|\\
\langle y_j|
\end{aligned}\right)D_j^{-1}(k_j)=0,
\end{aligned}
\end{equation}
then the regularization of $\psi_{j,x}\psi_j^{-1}$ at the point $k_j$ is realized.
Because the symmetries in (\ref{a3}) are valid for (\ref{d1}), we deduce that
$\psi_{j,x}\psi_j^{-1}$ is also analytic at the points $-k_j$ and $\pm\bar{k}_j$.
Similar arguments establish the regularity conditions of $\psi_{j,t}\psi_j^{-1}$.

Equation (\ref{d13}) and the analogous equation with respect to $t$ imply that the complex constants $\Upsilon_j$ and $\tilde\Upsilon_j$ exist:
\begin{equation}\label{d15}\begin{aligned}
\langle y_j|\psi_{j-1}(k_j)=\Upsilon_j,\\
\langle \tilde{y}_j|\psi_{j-1}(k_j)+\langle y_j|(\psi_{j-1})^\prime(k_j)=\tilde\Upsilon_j,
\end{aligned}
\end{equation}
which give (\ref{d4}). Moreover, the equations in (\ref{d14}) are equivalent to
\begin{equation}\label{d16}\begin{aligned}
\langle y_j|D_j^{-1}(k_j)=0,\\
\langle \tilde{y}_j|D_j^{-1}(k_j)+\langle y_j|(D_j^{-1})^\prime(k_j)=0.
\end{aligned}
\end{equation}
Substituting (\ref{d2}) and (\ref{d3}) into (\ref{d16}) and taking the Hermitian conjugate of the result equations yields
\begin{equation}\label{d17}
(|y_j\rangle,|\tilde{y}_j\rangle)+(|z_j\rangle,|\tilde{z}_j\rangle)M_j+\sigma_3(|z_j\rangle,|\tilde{z}_j\rangle)N_j=0,
\end{equation}
where
\begin{equation}\label{d18}\begin{aligned}
M_j=\left(K_-^\dagger(k_j)\left(\begin{array}{c}
\langle\tilde{y}_j|y_j\rangle\\
\langle y_j|y_j\rangle
\end{array}\right), (K_-^\dagger)^\prime(k_j)\left(\begin{array}{c}
\langle\tilde{y}_j|y_j\rangle\\
\langle y_j|y_j\rangle
\end{array}\right)+K_-^\dagger(k_j)\left(\begin{array}{c}
\langle\tilde{y}_j|\tilde{y}_j\rangle\\
\langle y_j|\tilde{y}_j\rangle
\end{array}\right)\right),\\
N_j=\left(K_+^\dagger(k_j)\left(\begin{array}{c}
\langle\tilde{y}_j|\sigma_3|y_j\rangle\\
\langle y_j|\sigma_3|y_j\rangle
\end{array}\right), (K_+^\dagger)^\prime(k_j)\left(\begin{array}{c}
\langle\tilde{y}_j|\sigma_3|y_j\rangle\\
\langle y_j|\sigma_3|y_j\rangle
\end{array}\right)+K_+^\dagger(k_j)\left(\begin{array}{c}
\langle\tilde{y}_j|\sigma_3|\tilde{y}_j\rangle\\
\langle y_j|\sigma_3|\tilde{y}_j\rangle
\end{array}\right)\right),
\end{aligned}
\end{equation}
and the matrix $K_\pm(k)$ is defined \cite{sam110-297} as
\begin{equation}\label{d19}
K_\pm(k)=\mp \frac{I}{k\pm\bar{k}_j}+\frac{\sigma_+^T}{(k\pm\bar{k}_j)^2}.
\end{equation}
Here, $\sigma_+^T$ denotes the transpose of $\sigma_+$ in (\ref{d7}).

Given $|y_j\rangle$ and $|\tilde{y}_j\rangle$, $|z_j\rangle$ and $|\tilde{z}_j\rangle$ can be solved from (\ref{d17}) as
\begin{equation}\label{d20}
(|z_j\rangle,|\tilde{z}_j\rangle)=-\frac{I+\sigma_3}{2}(|y_j\rangle,|\tilde{y}_j\rangle)(M_j+N_j)^{-1}
-\frac{I-\sigma_3}{2}(|y_j\rangle,|\tilde{y}_j\rangle)(M_j-N_j)^{-1},
\end{equation}
where $M_j\pm N_j$ takes the form 
\begin{equation}\label{d21}
M_j\pm N_j=\left(\begin{aligned}
&-\frac{a_{j,21}^\pm}{k_j^2-\bar{k}_j^2}+\frac{b_{j,11}^\pm}{(k_j^2-\bar{k}_j^2)^2}&\quad
-\frac{a_{j,22}^\pm}{k_j^2-\bar{k}_j^2}+\frac{b_{j,12}^\pm-b_{j,21}^\mp}{(k_j^2-\bar{k}_j^2)^2}+2\frac{c_{j,11}^\mp}{(k_j^2-\bar{k}_j^2)^3}\\
&-\frac{a_{j,11}^\pm}{k_j^2-\bar{k}_j^2}&-\frac{a_{j,12}^\pm}{k_j^2-\bar{k}_j^2}-\frac{b_{j,11}^\mp}{(k_j^2-\bar{k}_j^2)^2}
\end{aligned}\right),
\end{equation}
in terms of the definition (\ref{d9}).
It is easy to show that
\begin{equation}\label{d22}
\det(M_j\pm N_j)=\frac{\Delta_j^\pm}{(k_j^2-\bar{k}_j^2)^4},
\end{equation}
where $\Delta_j^\pm$ is defined by (\ref{d10}). Hence, equation (\ref{d20}) gives (\ref{d8})$-$(\ref{d10}).

This completes the regularization of the functions $\psi_{j,x}\psi^{-1}$ and $\psi_{j,t}\psi^{-1}$ at the point $k=k_j$.
The regularization of the functions $\psi_{j,x}\psi^{-1}$ and $\psi_{j,t}\psi^{-1}$ at the points $k=-k_j$ and $k=\pm\bar{k}_j$
can be discussed similarly. \hskip6cm \quad $\blacksquare$

It is convenient to rewrite $G(k)$ in (\ref{d1}) as
$$G(k)=X_j(k)D_j(k)T_j(k),$$
where $X_j(k)=D_N(k)\cdots D_{j+1}(k)$ and $T_j(k)=D_{j-1}(k)\cdots D_1(k)$ are analytic at $k=k_j$.
Then, near the point $k=k_j$, we have the expansion
\begin{equation}\label{d34}
\begin{aligned}
G(k)=&\frac{1}{(k-k_j)^2}X_j(k_j)\tilde{B}_jT_j(k_j)\\
&+\frac{1}{k-k_j}[X_j(k_j)B_jT_j(k_j)+X^\prime_j(k_j)\tilde{B}_jT_j(k_j)+X_j(k_j)\tilde{B}_jT^\prime_j(k_j)]+O(1),
\end{aligned}
\end{equation}
which implies, in view of (\ref{b2}), that
\begin{equation}\label{d35}
\tilde{A}_j=X_j(k_j)\tilde{B}_jT_j(k_j),
\end{equation}
and
\begin{equation}\label{d36}
A_j=X_j(k_j)B_jT_j(k_j)+X^\prime_j(k_j)\tilde{B}_jT_j(k_j)+X_j(k_j)\tilde{B}_jT^\prime_j(k_j),
\end{equation}
where
\begin{equation}\label{d37}
X^\prime_j(k_j)=\sum\limits_{l=j+1}^ND_N(k_j)\cdots D_{l+1}(k_j)D^\prime_l(k_j)D_{l-1}(k_j)\cdots D_{j+1}(k_j),
\end{equation}
and $T^\prime_j(k_j)$ has an equivalent definition. Because $\tilde{B}_j=|z_j\rangle\langle y_j|,
B_j=|\tilde{z}_j\rangle\langle y_j|+|z_j\rangle\langle \tilde{y}_j|$ and $\psi_{j-1}(k)=T_j(k)\psi_0(k)$, it follows that
\begin{equation}\label{d38}
\tilde{A}_j=X_j(k_j)\tilde{B}_j\psi_{j-1}(k_j)\psi_0^{-1}(k_j)=X_j(k_j)|z_j\rangle\Upsilon_j\psi_0^{-1}(k_j),
\end{equation}
in terms of (\ref{d15}). Similarly,
\begin{equation}\label{d39}
\begin{aligned}
A_j=&X_j(k_j)(|\tilde{z}_j\rangle\langle y_j|+|z_j\rangle\langle \tilde{y}_j|)\psi_{j-1}(k_j)\psi_0^{-1}(k_j)\\
&+X^\prime_j(k_j)|z_j\rangle\Upsilon_j\psi_0^{-1}(k_j)+X_j(k_j)|z_j\rangle\Upsilon_j\psi_{j-1}^{-1}(k_j)T^\prime_j(k_j)\\
=&X_j(k_j)|\tilde{z}_j\rangle\Upsilon_j\psi_0^{-1}(k_j)+X^\prime_j(k_j)|z_j\rangle\Upsilon_j\psi_0^{-1}(k_j)\\
&+X_j(k_j)|z_j\rangle[\tilde\Upsilon_j-\langle y_j|(\psi_{j-1})^\prime(k_j)]\psi_0^{-1}(k_j)\\
&+X_j(k_j)|z_j\rangle\Upsilon_j\psi_{j-1}^{-1}(k_j)[(\psi_{j-1})^\prime(k_j)\psi_0^{-1}(k_j)+\psi_{j-1}(k_j)(\psi_0^{-1})^\prime(k_j)],
\end{aligned}
\end{equation}
which reduces to
\begin{equation}\label{d40}
\begin{aligned}
A_j=&\left[X_j(k_j)|\tilde{z}_j\rangle+X^\prime_j(k_j)|z_j\rangle\right]\Upsilon_j\psi_0^{-1}(k_j)\\
&+X_j(k_j)|z_j\rangle\left[\tilde\Upsilon_j\psi_0^{-1}(k_j)+\Upsilon_j(\psi_0^{-1})^\prime(k_j)\right].
\end{aligned}
\end{equation}
Equation (\ref{c4}) implies that the solution of the GI equation is still given by (\ref{b12}), but the form of $A_j$ is different.
As an example, we shall consider the case of $N=1$, that is, the case of the two-order soliton solution.

It is noted that, for $N=1$, $G=D_1$, and $A_1=B_1$, the two-order soliton can be derived directly.
In this case, The zero seed solution and associated eigenfunction $\psi_0$ are chosen as disscused in Section 3.
Because
\begin{equation}\label{d27}
\psi_0^{-1}(k)={\rm e}^{i\theta(k)\sigma_3},\quad (\psi_0^{-1})^\prime(k)=i\theta^\prime(k){\rm e}^{i\theta(k)\sigma_3},
\end{equation}
the vectors in (\ref{d4}) can be written as
\begin{equation}\label{d23}
\langle y_1|=\left({\rm e}^{\theta_1},{\rm e}^{-\theta_1}\right),\quad \langle \tilde{y}_1|=\left(\alpha_1{\rm e}^{\theta_1},\alpha_2{\rm e}^{-\theta_1}\right),
\end{equation}
where $\alpha_1,\alpha_2$ and $\theta_1$ are defined as
\begin{equation}\label{d24}
{\rm e}^{\theta_1}=\beta_1{\rm e}^{i\theta(k_1)},\quad \alpha_1=\varepsilon_1+i\theta^\prime(k_1),\quad \alpha_2=\varepsilon_2-i\theta^\prime(k_1),
\end{equation}
in terms of the chosen constant vectors  $\Upsilon_1=(\beta_1,\beta_1^{-1})$ and $\tilde\Upsilon_1=(\varepsilon_1\beta_1,\varepsilon_2\beta_1^{-1})$.
Furthermore, if we let $k_1=\xi_1+i\eta_1$ and $\theta_1=-r_1+i\phi_1$, then
\begin{equation}\label{d25}
\begin{aligned}
r_1&=2\xi_1\eta_1x+8\xi_1\eta_1(\xi_1^2-\eta_1^2)t+r_0,\\
\phi_1&=(\xi_1^2-\eta_1^2)x+2[(\xi_1^2-\eta_1^2)^2-4\xi_1^2\eta_1^2]t+\phi_0,
\end{aligned}
\end{equation}
where $r_0$ and $\phi_0$ are real constants defined by $\beta_1=e^{-r_0+i\phi_0}$.
For the two-order solution of the GI equation, we find, from (\ref{d6}) for $j=1$ (omitted here for convenience), that
\begin{equation}\label{d26}
\begin{aligned}
a_{11}^+&=2\left(k_1{\rm e}^{-2r_1}+\bar{k}_1{\rm e}^{2r_1}\right),\\
a_{12}^+&=2\left(k_1\bar\alpha_1{\rm e}^{-2r_1}+\bar{k}_1\bar\alpha_2{\rm e}^{2r_1}\right),\\
a_{21}^+&=2\left(k_1\alpha_1{\rm e}^{-2r_1}+\bar{k}_1\alpha_2{\rm e}^{2r_1}\right),\\
a_{22}^+&=2\left(k_1\alpha_1\bar\alpha_1{\rm e}^{-2r_1}+\bar{k}_1\alpha_2\bar\alpha_2{\rm e}^{2r_1}\right),\\
b_{11}^+&=2\left((k_1^2+\bar{k}_1^2){\rm e}^{-2r_1}+2k_1\bar{k}_1{\rm e}^{2r_1}\right),\\
b_{11}^-&=2\left(2k_1\bar{k}_1{\rm e}^{-2r_1}+(k_1^2+\bar{k}_1^2){\rm e}^{2r_1}\right),\\
b_{12}^+&=2\left((k_1^2+\bar{k}_1^2)\bar\alpha_1{\rm e}^{-2r_1}+2k_1\bar{k}_1\bar\alpha_2{\rm e}^{2r_1}\right),\\
b_{21}^-&=2\left(2k_1\bar{k}_1\alpha_1{\rm e}^{-2r_1}+(k_1^2+\bar{k}_1^2)\alpha_2{\rm e}^{2r_1}\right),\\
c_{11}^-&=2\left((\bar{k}_1^3+3\bar{k}_1k_1^2){\rm e}^{-2r_1}+(k_1^3+3k_1\bar{k}_1^2){\rm e}^{2r_1}\right).\\
\end{aligned}
\end{equation}
With these representations in hand, we find that
\begin{equation}\label{f6}
\begin{aligned}
\Delta^+=&(k_1^2-\bar{k}_1^2)^2(a_{12}^+a_{21}^+-a_{11}^+a_{22}^+)+2a_{11}^+c_{11}^--b_{11}^+b_{11}^-\\
&+(k_1^2-\bar{k}_1^2)[(a_{21}^+b_{11}^--a_{11}^+b_{21}^-)+(a_{11}^+b_{12}^+-a_{12}^+b_{11}^+)]\\
=&-4(k_1^2-\bar{k}_1^2)^2|k_1|^2|\alpha_2-\alpha_1|^2+16k\bar{k}_1(k_1e^{-r_1}+\bar{k}_1e^{r_1})^2\\
&+4(k_1^2-\bar{k}_1^2)^2-4(k_1^2-\bar{k}_1^2)^2[k_1(\alpha_2-\alpha_1)-\bar{k}_1(\bar\alpha_2-\bar\alpha_1)]\\
=&4(k_1^2-\bar{k}_1^2)^2[1-k_1(\alpha_2-\alpha_1)][1+\bar{k}_1(\bar\alpha_2-\bar\alpha_1)]\\
&+16k\bar{k}_1(k_1e^{-r_1}+\bar{k}_1e^{r_1})^2.
\end{aligned}
\end{equation}
In addition, from (\ref{d7})$-$(\ref{d9}), we know that
\begin{equation}\label{f7}
\begin{aligned}
(B_1)_{12}=&e^{-i\phi_1}[\bar\alpha_1\alpha_2\Lambda_1+\bar\alpha_1\tilde\Lambda_1+\alpha_2\Omega_1+\tilde\Omega_1]\\
=&\frac{e^{-i\phi_1}}{\Delta^+}[(k_1^2-\bar{k}_1^2)^3(-\bar\alpha_1\alpha_2a_{11}^++\bar\alpha_1a_{21}^++\alpha_2a_{12}^+-a_{22}^+)\\
&\qquad+(k_1^2-\bar{k}_1^2)^2(-\bar\alpha_1b_{11}^++\alpha_2b_{11}^-+b_{12}^+-b_{21}^-)+2(k_1^2-\bar{k}_1^2)c_{11}^-]\\
=&\frac{e^{-i\phi_1}}{\Delta^+}\left\{4(k_1^2-\bar{k}_1^2)^2|k_1|^2[(\alpha_2-\alpha_1)e^{-r_1}+(\bar\alpha_2-\bar\alpha_1)e^{r_1}]\right.\\
&\qquad+\left.4(k_1^2-\bar{k}_1^2)[(\bar{k}_1^3+3\bar{k}_1k_1^2)e^{-r_1}+(k_1^3+3k_1\bar{k}_1^2)e^{r_1}]\right\}.
\end{aligned}
\end{equation}
Note that $k_1=\xi_1+i\eta_1$, and, by taking $\rho:=\alpha_2-\alpha_1$, we have
\begin{equation}\label{f8}
\begin{aligned}
\rho=-2i\theta^\prime(k_1)+\varepsilon_2-\varepsilon_1=\mu-i\nu,\\
\mu=4\eta_1[x+4(3\xi_1^2-\eta_1^2)t+\epsilon_1],\\
\nu=4\xi_1[x+4(\xi_1^2-3\eta_1^2)t+\epsilon_2].
\end{aligned}
\end{equation}

Hence, The two-order soliton solution takes the form
\begin{equation}\label{f9}
q=4{\rm e}^{-i\tilde\phi_1}\frac{N_2}{D_2},
\end{equation}
where
\begin{equation}\label{f10}
\begin{aligned}
&N_2=4|k_1|^2\left(k_1{\rm e}^{-r_1}+\bar{k}_1{\rm e}^{r_1}\right)^2+(k_1^2-\bar{k}_1^2)^2(1-k_1\rho)(1+\bar{k}_1\bar\rho),\\
&D_2=w{\rm e}^{-r_1}-\bar{w}{\rm e}^{r_1}, \quad w=(k_1^2-\bar{k}_1^2)^3|k_1|^2\rho+(k_1^2-\bar{k}_1^2)(\bar{k}_1^3+3\bar{k}_1k_1^2),
\end{aligned}
\end{equation}
and $\tilde\phi_1=\phi_1-\pi/2$. Here $r_1$ and $\phi_1$ are defined according to (\ref{d25}). The Figure 1 illustrates the two-order soliton.\\
\begin{figure}[h]
\centering
\includegraphics[width=5cm,angle=-90]{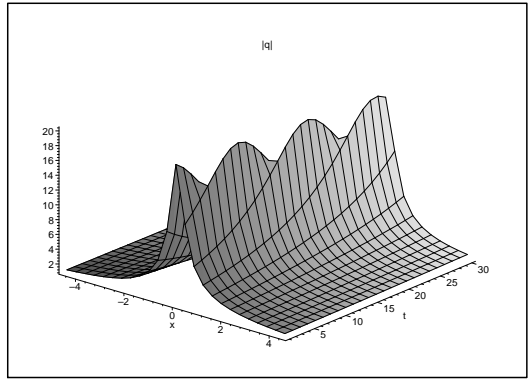}
\caption{Solution (\ref{f9}) with $k_1=0.01+0.1i, \beta_1=1, \varepsilon_1=\varepsilon_2$.}
\end{figure}

\setcounter{equation}{0}
\section{Discussions}
We note that the relationship between $\tilde{B}_j$ in (\ref{d7}) and $D_j^{-1}(k_j)$ in (\ref{d3}) can be constructed (see the Appendix) as
\begin{equation}\label{g1}
\sigma_2\tilde{B}_j^T\sigma_2=\frac{(k_j^2-\bar{k}_j^2)^2}{4k_j^2}D_j^{-1}(k_j),
\end{equation}
which is similar to the equation given in (\ref{b26}). However, the relationship between $B_j$ and $D_j^{-1}(k_j)$ is very complicated,
such that it is difficult to establish the relationship between $D_j(k)$ and $D_j^{-1}(k)$. Hence, it is difficult by this method to obtain
the explicit form of higher-order ($N>2$) soliton to the GI equation. This problem remains open.

\section*{Acknowledgments}

This work was supported by the National Natural Science Foundation of China (Projects 11471295).
We thank LetPub for its linguistic assistance during the preparation of this manuscript.

\setcounter{equation}{0}
\section{Appendix}
In this Appendix, we derive the relation (\ref{g1}). To this end, we introduce a set of Lemmas.

\begin{lemma}\label{l1}
Let the row vectors $\langle y_j|,\langle\tilde{y}_j|, (j=1,2,\cdots,N)$ and the scalars $a_{il}^\pm, b_{il}^\pm, c_{11}^\pm,$ $(i,l=1,2)$ are define by Proposition \ref{p3}.
Then these scalars $a_{il}^\pm,b_{il}^\pm,c_{11}^\pm$ satisfy the following symmetries
\begin{equation}\label{A1}
\begin{aligned}
\overline{a_{11}^\pm}=a_{11}^\mp,\quad\overline{a_{12}^\pm}=a_{21}^\mp, \quad \overline{a_{22}^\pm}=a_{22}^\mp,\\
\overline{b_{11}^\pm}=b_{11}^\pm,\quad\overline{b_{12}^\pm}=b_{21}^\pm,\quad \overline{c_{11}^\pm}=c_{11}^\mp.
\end{aligned}
\end{equation}
Moreover, we have
\begin{equation}\label{A2}\begin{aligned}
\overline{\Delta^\pm}=\Delta^\mp ,\quad \zeta_jb_{11}^++\bar\zeta_jb_{11}^-=c_{11}^+,
\end{aligned}
\end{equation}
\begin{equation}\label{A3}\begin{aligned}
\frac{I\pm\sigma_3}{2}2\zeta_j[a_{11}^\pm|\tilde{y}_j\rangle-a_{12}^\pm|y_j\rangle]=\frac{I\pm\sigma_3}{2}[b_{11}^\pm|\tilde{y}_j\rangle-b_{12}^\pm|y_j\rangle],
\end{aligned}
\end{equation}
\begin{equation}\label{A4}\begin{aligned}
&\frac{I\pm\sigma_3}{2}[a_{11}^\pm|\tilde{y}_j\rangle-a_{12}^\pm|y_j\rangle]\langle\tilde{y}_j|\frac{I\mp\sigma_3}{2}\\
&\quad=\frac{I\pm\sigma_3}{2}[a_{21}^\pm|\tilde{y}_j\rangle-a_{22}^\pm|y_j\rangle]\langle y_j|\frac{I\mp\sigma_3}{2},\\
\end{aligned}
\end{equation}
or equivalently
\begin{equation}\label{A5}\begin{aligned}
\qquad \bar\zeta_jb_{11}^++\zeta_jb_{11}^-=&c_{11}^-,\\
\end{aligned}
\end{equation}
\begin{equation}\label{A6}\begin{aligned}
2\bar\zeta_j[\langle\tilde{y}_j|\overline{a_{11}^\pm}-\langle y_j|\overline{a_{12}^\pm}]\frac{I\pm\sigma_3}{2}
=&[\langle\tilde{y}_j|\overline{b_{11}^\pm}-\langle y_j|\overline{b_{12}^\pm}]\frac{I\pm\sigma_3}{2},
\end{aligned}
\end{equation}
\begin{equation}\label{A7}\begin{aligned}
&\frac{I\mp\sigma_3}{2}|\tilde{y}_j\rangle[\langle\tilde{y}_j|\overline{a_{11}^\pm}-\langle y_j|\overline{a_{12}^\pm}]\frac{I\pm\sigma_3}{2}\\
&\quad=\frac{I\mp\sigma_3}{2}|y_j\rangle[\langle\tilde{y}_j|\overline{a_{21}^\pm}-\langle y_j|\overline{a_{22}^\pm}]\frac{I\pm\sigma_3}{2}.\\
\end{aligned}
\end{equation}
\end{lemma}
Proof: The proof of (\ref{A1}) and (\ref{A2}) can be finished by directly calculation from the definition (\ref{d6}).
We proceed to prove the first case of (\ref{A3}). For algorithmic convenience, we take the column vectors $|y_j\rangle, |\tilde{y}_j\rangle$ in the form
$$|y_j\rangle=(f_1,f_2)^T,\quad |\tilde{y}_j\rangle=(g_1, g_2)^T,$$
then
$$\langle y_j|=(\bar{f}_1,\bar{f}_2),\quad \langle\tilde{y}_j|=(\bar{g}_1, \bar{g}_2).$$
Here, we note, for the column vector $(a_+,a_-)^T$, that
$$\frac{I+\sigma_3}{2}\left(\begin{array}{c}
a_+\\
a_-
\end{array}\right)=\frac{I+\sigma_3}{2}\left(\begin{array}{c}
a_+\\
*
\end{array}\right),\quad \frac{I-\sigma_3}{2}\left(\begin{array}{c}
a_+\\
a_-
\end{array}\right)=\frac{I-\sigma_3}{2}\left(\begin{array}{c}
*\\
a_-
\end{array}\right),$$
where the star denotes an arbitrary element and can be chosen as zero in particularly.
On use of the definition (\ref{d6}), we have
$$\begin{aligned}
&\frac{I+\sigma_3}{2}2\zeta_j[a_{11}^+|\tilde{y}_j\rangle-a_{12}^+|y_j\rangle]
=2\zeta_j\frac{I+\sigma_3}{2}\left(\begin{array}{c}
a_{11}^+g_1-a_{12}^+f_1\\
*
\end{array}\right)\\
&=4\zeta_j\frac{I+\sigma_3}{2}\left(\begin{array}{c}
(\zeta_jf_1\bar{f_1}+\bar\zeta_jf_2\bar{f_2})g_1-(\zeta_jg_1\bar{f_1}+\bar\zeta_jg_2\bar{f_2})f_1\\
*
\end{array}\right)\\
&=4\zeta_j\bar\zeta_j\frac{I+\sigma_3}{2}\left(\begin{array}{c}
\bar{f_2}(f_2g_1-f_1g_2)\\
*
\end{array}\right),
\end{aligned}$$
and
$$\begin{aligned}
&\frac{I+\sigma_3}{2}[b_{11}^+|\tilde{y}_j\rangle-b_{12}^+|y_j\rangle]\\
&=2\frac{I+\sigma_3}{2}\left(\begin{array}{c}
[(\zeta_j^2+\bar\zeta_j^2)f_1\bar{f_1}+2\zeta_j\bar\zeta_jf_2\bar{f_2}]g_1-[(\zeta_j^2+\bar\zeta_j^2)g_1\bar{f_1}+2\zeta_j\bar\zeta_jg_2\bar{f_2}]f_1\\
*
\end{array}\right)\\
&=4\zeta_j\bar\zeta_j\frac{I+\sigma_3}{2}\left(\begin{array}{c}
\bar{f_2}(f_2g_1-f_1g_2)\\
*
\end{array}\right).
\end{aligned}$$
Then the first case of (\ref{A3}) is proved, and the second case can be proved similarly.

For the first case of (\ref{A4}), we note that
$$\begin{aligned}
&\frac{I+\sigma_3}{2}[a_{11}^+|\tilde{y}_j\rangle-a_{12}^+|y_j\rangle]\langle\tilde{y}_j|\frac{I-\sigma_3}{2}\\
&=2\frac{I+\sigma_3}{2}\left(\begin{array}{c}
\bar\zeta_j\bar{f_2}(f_2g_1-f_1g_2)\\
*
\end{array}\right)(*\quad \bar{g_2})\frac{I-\sigma_3}{2}\\
&=2\frac{I+\sigma_3}{2}\left(\begin{matrix}
*&\bar\zeta_j\bar{f_2}\bar{g_2}(f_2g_1-f_1g_2)\\
*&*
\end{matrix}\right)\frac{I-\sigma_3}{2},
\end{aligned}$$
and
$$\begin{aligned}
&\frac{I+\sigma_3}{2}[a_{21}^+|\tilde{y}_j\rangle-a_{22}^+|y_j\rangle]\langle y_j|\frac{I-\sigma_3}{2}\\
&=2\frac{I+\sigma_3}{2}\left(\begin{array}{c}
\bar\zeta_j\bar{g_2}(f_2g_1-f_1g_2)\\
*
\end{array}\right)(*\quad \bar{f_2})\frac{I-\sigma_3}{2}\\
&=2\frac{I+\sigma_3}{2}\left(\begin{matrix}
*&\bar\zeta_j\bar{f_2}\bar{g_2}(f_2g_1-f_1g_2)\\
*&*
\end{matrix}\right)\frac{I-\sigma_3}{2}.
\end{aligned}$$
This completes the proof of the first case of (\ref{A4}), and the proof of the second case can be fulfilled similarly. \qquad$\blacksquare$

\begin{lemma}\label{l2}
Let the row vectors $\langle y_j|,\langle\tilde{y}_j|, (j=1,2,\cdots,N)$ and the scalars $a_{il}^\pm, b_{il}^\pm, c_{11}^\pm, (i,l=1,2)$ are define by Proposition \ref{p3}.
Then
\begin{equation}\label{d28}
\begin{aligned}
&\frac{I\pm\sigma_3}{2}\left\{(a_{12}^\mp a_{21}^\mp-a_{11}^\mp a_{22}^\mp)+2\bar\zeta_j\left[
|\tilde{y}_j\rangle\left(\langle\tilde{y}_j|\overline{a_{11}^\pm}-\langle y_j|\overline{a_{12}^\pm}\right)\right.\right.\\
&\qquad\left.\left.-|y_j\rangle\left(\langle\tilde{y}_j|\overline{a_{21}^\pm}-\langle y_j|\overline{a_{22}^\pm}\right)\right]\right\}\frac{I\pm\sigma_3}{2}=\textbf{0},
\end{aligned}
\end{equation}
\begin{equation}\label{d29}
\begin{aligned}
&\frac{I\pm\sigma_3}{2}\left\{[a_{21}^\mp b_{11}^\pm-a_{12}^\mp b_{11}^\mp+a_{11}^\mp(b_{12}^\mp-b_{21}^\pm)]
+2\bar\zeta_j\left(b_{11}^\mp|\tilde{y}_j\rangle-b_{12}^\mp|y_j\rangle\right)\langle y_j|\right.\\
&\quad\left.+2|y_j\rangle\left[(\zeta_j^2+\bar\zeta_j^2)\left(\langle\tilde{y}_j|\overline{a_{11}^\pm}-\langle y_j|\overline{a_{12}^\pm}\right)
-\bar\zeta_j\left(\langle\tilde{y}_j|\overline{b_{11}^\pm}-\langle y_j|\overline{b_{12}^\pm}\right)\right]\right\}\frac{I\pm\sigma_3}{2}\\
&\qquad=-4\zeta_j^2\sigma_2\frac{I\mp\sigma_3}{2}\{\left(a_{11}^\mp|\tilde{y}_j\rangle-a_{12}^\mp|y_j\rangle\right)\langle y_j|\}\frac{I\mp\sigma_3}{2}\sigma_2,
\end{aligned}
\end{equation}
\begin{equation}\label{d30}
\begin{aligned}
&\frac{I\pm\sigma_3}{2}\left\{(2a_{11}^\mp c_{11}^\pm-b_{11}^\mp b_{11}^\pm)
+\left[2(\zeta_j^2+\bar\zeta_j^2)\overline{b_{11}^\mp}-4\bar\zeta_j\overline{c_{11}^\mp}\right]|y_j\rangle\langle y_j|\right\}\frac{I\pm\sigma_3}{2}\\
&\qquad=4\zeta_j^2\sigma_2\frac{I\mp\sigma_3}{2}\left[b_{11}^\pm|y_j\rangle\langle y_j|\right]\frac{I\mp\sigma_3}{2}\sigma_2.
\end{aligned}
\end{equation}
\end{lemma}
Proof: 
To prove the first case of (\ref{d29}), we note from the definition (\ref{d6}) that
$$\begin{aligned}
&a_{21}^- b_{11}^+-a_{12}^- b_{11}^-+a_{11}^-(b_{12}^--b_{21}^+)\\
&=4(\zeta_j^2-\bar\zeta_j^2)\left[\zeta_jf_1f_2\overline{(f_1g_2-f_2g_1)}-\overline{\zeta_jf_1f_2}(f_1g_2-f_2g_1)\right],
\end{aligned}$$
and
$$\begin{aligned}
&\frac{I+\sigma_3}{2}\left\{2\bar\zeta_j\left(b_{11}^-|\tilde{y}_j\rangle-b_{12}^-|y_j\rangle\right)\langle y_j|\right.\\
&\quad\left.+2|y_j\rangle\left[(\zeta_j^2+\bar\zeta_j^2)\left(\langle\tilde{y}_j|\overline{a_{11}^+}-\langle y_j|\overline{a_{12}^+}\right)
-\bar\zeta_j\left(\langle\tilde{y}_j|\overline{b_{11}^+}-\langle y_j|\overline{b_{12}^+}\right)\right]\right\}\frac{I+\sigma_3}{2}\\
&=\frac{I+\sigma_3}{2}\left\{2\bar\zeta_j\left(b_{11}^-g_1-b_{12}^-f_1\right)\bar{f_1}\right.\\
&\qquad\quad\left.+2f_1\left[(\zeta_j^2+\bar\zeta_j^2)\left(\overline{a_{11}^+g_1-a_{12}^+f_1}\right)
-\bar\zeta_j\left(\overline{b_{11}^+g_1-b_{12}^+f_1}\right)\right]\right\}\frac{I+\sigma_3}{2}\\
&=\frac{I+\sigma_3}{2}\left\{-4(\zeta_j^2-\bar\zeta_j^2)\left[\bar\zeta_j\left(\zeta_j^2+\bar\zeta_j^2\right)\overline{f_1f_2}(f_1g_2-f_2g_1)\right.\right.\\
&\qquad\quad\left.+\left.\zeta_j f_1f_2\overline{(f_1g_2-f_2g_1)}\right]\right\}\frac{I+\sigma_3}{2}.\\
\end{aligned}$$
Then the left-hand side of (\ref{d29}) in the first case reduces to
$$\begin{aligned}
&\frac{I+\sigma_3}{2}\left\{-4\zeta_j^2[2\overline{\zeta_jf_1}(f_1g_2-f_2g_1)]\overline{f_2}\right\}\frac{I+\sigma_3}{2}\\
&=\frac{I+\sigma_3}{2}\left\{-4\zeta_j^2(a_{11}^-g_2-a_{12}^-f_2)\overline{f_2}\right\}\frac{I+\sigma_3}{2}\\
&=-4\zeta_j^2\sigma_2\frac{I-\sigma_3}{2}\{\left(a_{11}^-|\tilde{y}_j\rangle-a_{12}^-|y_j\rangle\right)\langle y_j|\}\frac{I-\sigma_3}{2}\sigma_2.
\end{aligned}$$
This finished the prove of the first case of (\ref{d29}). The prove for the second case of (\ref{d29}), as well as for (\ref{d28}) and (\ref{d30}) are similar. \qquad$\blacksquare$

\begin{lemma}\label{l3}
Let the row vectors $\langle y_j|,\langle\tilde{y}_j|$ and the column vectors $|z_j\rangle,|\tilde{z}_j\rangle$  are defined by Proposition\ref{p3}, then
\begin{equation}\label{d31}\begin{aligned}
\frac{I\pm\sigma_3}{2}&\left[\frac{4\zeta_j\bar\zeta_j}{(\zeta_j^2-\bar\zeta_j^2)^2}|y_j\rangle\langle z_j|
+\frac{2\zeta_j}{\zeta_j^2-\bar\zeta_j^2}[|\tilde{y}_j\rangle\langle z_j|+|y_j\rangle\langle \tilde{z}_j|]\right]\frac{I\mp\sigma_3}{2}\\
&=-\frac{4\zeta_j^2}{(\zeta_j^2-\bar\zeta_j^2)^2}\frac{I\pm\sigma_3}{2}|z_j\rangle\langle y_j|\frac{I\mp\sigma_3}{2},
\end{aligned}
\end{equation}
and
\begin{equation}\label{d32}\begin{aligned}
&\frac{I\pm\sigma_3}{2}\left[I+\frac{2(\zeta_j^2+\bar\zeta_j^2)}{(\zeta_j^2-\bar\zeta_j^2)^2}|y_j\rangle\langle z_j|
+\frac{2\bar\zeta_j}{\zeta_j^2-\bar\zeta_j^2}[|\tilde{y}_j\rangle\langle z_j|+|y_j\rangle\langle \tilde{z}_j|]\right]\frac{I\pm\sigma_3}{2}\\
&\qquad=\frac{4\zeta_j^2}{(\zeta_j^2-\bar\zeta_j^2)^2}\sigma_2\frac{I\mp\sigma_3}{2}|z_j\rangle\langle y_j|\frac{I\mp\sigma_3}{2}\sigma_2.
\end{aligned}
\end{equation}
\end{lemma}
Proof: We shall prove the first case of (\ref{d31}) by virtue of the definition (\ref{d8}) and Lemma \ref{l1}. It is easy to show that
\begin{equation}\label{d33}
\begin{aligned}
\langle y_j|\Lambda\frac{I-\sigma_3}{2}=\langle y_j|\Lambda_2\frac{I-\sigma_3}{2},\quad
\langle y_j|\Lambda\frac{I+\sigma_3}{2}=\langle y_j|\Lambda_1\frac{I+\sigma_3}{2}.
\end{aligned}
\end{equation}
With these identities in hand and on use of the definition (\ref{d8}), we have
$$\begin{aligned}
&\frac{I+\sigma_3}{2}\left[\frac{4\zeta_j\bar\zeta_j}{(\zeta_j^2-\bar\zeta_j^2)^2}|y_j\rangle\langle z_j|
+\frac{2\zeta_j}{\zeta_j^2-\bar\zeta_j^2}[|\tilde{y}_j\rangle\langle z_j|+|y_j\rangle\langle \tilde{z}_j|]\right]\frac{I-\sigma_3}{2}\\
&=\frac{I+\sigma_3}{2}\left\{\frac{4\zeta_j\bar\zeta_j}{(\zeta_j^2-\bar\zeta_j^2)^2}|y_j\rangle\left(\langle\tilde{y}_j|\bar\Lambda+\langle y_j|\bar\Omega\right)\right.\\
&\quad\left.+\frac{2\zeta_j}{\zeta_j^2-\bar\zeta_j^2}\left[|\tilde{y}_j\rangle\left(\langle\tilde{y}_j|\bar\Lambda+\langle y_j|\bar\Omega\right)
+|y_j\rangle\left(\langle\tilde{y}_j|\bar{\tilde\Lambda}+\langle y_j|\bar{\tilde\Omega}\right)\right]\right\}\frac{I-\sigma_3}{2}\\
&=\frac{I+\sigma_3}{2}\left\{\frac{4\zeta_j\bar\zeta_j}{(\zeta_j^2-\bar\zeta_j^2)^2}|y_j\rangle\left(\langle\tilde{y}_j|\bar\Lambda_2+\langle y_j|\bar\Omega_2\right)\right.\\
&\quad\left.+\frac{2\zeta_j}{\zeta_j^2-\bar\zeta_j^2}\left[|\tilde{y}_j\rangle\left(\langle\tilde{y}_j|\bar\Lambda_2+\langle y_j|\bar\Omega_2\right)
+|y_j\rangle\left(\langle\tilde{y}_j|\bar{\tilde\Lambda}_2+\langle y_j|\bar{\tilde\Omega}_2\right)\right]\right\}\frac{I-\sigma_3}{2}.\\
\end{aligned}
$$
Substitution of (\ref{d9}) into the right-hand side of above equation implies that its right-hand side reduces to
$$\begin{aligned}
&\frac{I+\sigma_3}{2\Delta^+}\left\{2\zeta_j(\zeta_j^2-\bar\zeta_j^2)^2
\left[|\tilde{y}_j\rangle\left(\langle\tilde{y}_j|\overline{a_{11}^-}-\langle y_j|\overline{a_{12}^-}\right)
-|y_j\rangle\left(\langle\tilde{y}_j|\overline{a_{21}^-}-\langle y_j|\overline{a_{22}^-}\right)\right]\right.\\
&\quad+2\zeta_j(\zeta_j^2-\bar\zeta_j^2)|y_j\rangle\left[2\bar\zeta_j\left(\langle\tilde{y}_j|\overline{a_{11}^-}-\langle y_j|\overline{a_{12}^-}\right)
-\left(\langle\tilde{y}_j|\overline{b_{11}^-}-\langle y_j|\overline{b_{12}^-}\right)\right]\\
&\quad\left.+2\zeta_j(\zeta_j^2-\bar\zeta_j^2)\left(\overline{b_{11}^+}|\tilde{y}_j\rangle-\overline{b_{21}^+}|y_j\rangle\right)\langle y_j|
+4\zeta_j\left(\bar\zeta_j\overline{b_{11}^+}-\overline{c_{11}^+}\right)|y_j\rangle\langle y_j|\right\}\frac{I-\sigma_3}{2}\\
&=\frac{I+\sigma_3}{2\Delta^+}\left[2\zeta_j(\zeta_j^2-\bar\zeta_j^2)\left(b_{11}^+|\tilde{y}_j\rangle-b_{12}^+|y_j\rangle\right)\langle y_j|
-4\zeta_j^2b_{11}^-|y_j\rangle\langle y_j|\right]\frac{I-\sigma_3}{2}\\
&=\frac{I+\sigma_3}{2\Delta^+}\left[4\zeta_j^2(\zeta_j^2-\bar\zeta_j^2)\left(a_{11}^+|\tilde{y}_j\rangle-a_{12}^+|y_j\rangle\right)\langle y_j|
-4\zeta_j^2b_{11}^-|y_j\rangle\langle y_j|\right]\frac{I-\sigma_3}{2}\\
&=-\frac{4\zeta_j^2}{(\zeta_j^2-\bar\zeta_j^2)^2}\frac{I+\sigma_3}{2}\left(\Lambda_1|\tilde{y}_j\rangle+\Omega_1|y_j\rangle\right)\langle y_j|\frac{I-\sigma_3}{2},\\
\end{aligned}
$$
in terms of Lemma \ref{l1}. Thus we have
$$\begin{aligned}
&\frac{I+\sigma_3}{2}\left[\frac{4\zeta_j\bar\zeta_j}{(\zeta_j^2-\bar\zeta_j^2)^2}|y_j\rangle\langle z_j|
+\frac{2\zeta_j}{\zeta_j^2-\bar\zeta_j^2}[|\tilde{y}_j\rangle\langle z_j|+|y_j\rangle\langle \tilde{z}_j|]\right]\frac{I-\sigma_3}{2}\\
&=-\frac{4\zeta_j^2}{(\zeta_j^2-\bar\zeta_j^2)^2}\frac{I+\sigma_3}{2}\left(\Lambda|\tilde{y}_j\rangle+\Omega|y_j\rangle\right)\langle y_j|\frac{I-\sigma_3}{2}\\
&=-\frac{4\zeta_j^2}{(\zeta_j^2-\bar\zeta_j^2)^2}\frac{I+\sigma_3}{2}|z_j\rangle\langle y_j|\frac{I-\sigma_3}{2}.\\
\end{aligned}$$
This complete the prove of the first case of (\ref{d31}). The second case of (\ref{d31}) can be proved similarly.

We are now turning to the proof of (\ref{d32}). For the first case, by using the definition (\ref{d8}), we find that the left-hand side of (\ref{d32}) becomes
$$
\begin{aligned}
&\frac{I+\sigma_3}{2}\left\{I+\frac{2(\zeta_j^2+\bar\zeta_j^2)}{(\zeta_j^2-\bar\zeta_j^2)^2}|y_j\rangle\left(\langle\tilde{y}_j|\bar\Lambda_1+\langle y_j|\bar\Omega_1\right)\right.\\
&\qquad\left.-\frac{2\zeta_j}{\zeta_j^2-\bar\zeta_j^2}\left[|\tilde{y}_j\rangle\left(\langle\tilde{y}_j|\bar\Lambda_1+\langle y_j|\bar\Omega_1\right)
+|y_j\rangle\left(\langle\tilde{y}_j|\bar{\tilde{\Lambda}}_1+\langle y_j|\bar{\tilde{\Omega}}_1\right)\right]\right\}\frac{I+\sigma_3}{2}\\
\end{aligned}
$$
which, by using representations (\ref{d9}) and (\ref{d10}), gives
$$
\begin{aligned}
&\frac{I+\sigma_3}{2\Delta^-}\left\{(\zeta_j^2-\bar\zeta_j^2)^2\left[(a_{21}^-a_{12}^--a_{11}^-a_{22}^-)\right.\right.\\
&\qquad\left.+2\bar\zeta_j\left[|\tilde{y}_j\rangle\left(\langle\tilde{y}_j|\overline{a_{11}^+}-\langle y_j|\overline{a_{12}^+}\right)
-|y_j\rangle\left(\langle\tilde{y}_j|\overline{a_{21}^+}-\langle y_j|\overline{a_{22}^+}\right)\right]\right]\\
&+(\zeta_j^2-\bar\zeta_j^2)\left[a_{21}^-b_{11}^+-a_{12}^-b_{11}^-+a_{11}^-(b_{12}^--b_{21}^+)
+2\bar\zeta_j\left(b_{11}^-|\tilde{y}_j\rangle-b_{12}^-|y_j\rangle\right)\langle y_j|\right.\\
&\qquad\left.+2|y_j\rangle\left[(\zeta_j^2+\bar\zeta_j^2)\left(\langle\tilde{y}_j|\overline{a_{11}^+}-\langle y_j|\overline{a_{12}^+}\right)
-\bar\zeta_j\left(\langle\tilde{y}_j|\overline{b_{11}^+}-\langle y_j|\overline{b_{12}^+}\right)\right]\right]\\
&\left.+(2a_{11}^- c_{11}^+-b_{11}^- b_{11}^+)+\left[2(\zeta_j^2+\bar\zeta_j^2)\overline{b_{11}^-}-4\bar\zeta_j\overline{c_{11}^-}\right]\right\}\frac{I+\sigma_3}{2}.
\end{aligned}
$$
Now according to the first case of Lemma \ref{l2}, we find that
$$\begin{aligned}
&\frac{I+\sigma_3}{2}\left\{I+\frac{2(\zeta_j^2+\bar\zeta_j^2)}{(\zeta_j^2-\bar\zeta_j^2)^2}|y_j\rangle\langle z_j|
+\frac{2\bar\zeta_j}{\zeta_j^2-\bar\zeta_j^2}[|\tilde{y}_j\rangle\langle z_j|+|y_j\rangle\langle \tilde{z}_j|]\right\}\frac{I+\sigma_3}{2}\\
&=\frac{4\zeta_j^2}{\Delta^-}\sigma_2\frac{I-\sigma_3}{2}
\left[-(\zeta_j^2-\bar\zeta_j^2)(a_{11}^-|\tilde{y}_j\rangle-a_{12}^-|y_j\rangle)+b_{11}^+|y_j\rangle\right]\langle y_j|\frac{I-\sigma_3}{2}\sigma_2\\
&=\frac{4\zeta_j^2}{(\zeta_j^2-\bar\zeta_j^2)^2}\sigma_2\frac{I-\sigma_3}{2}(\Lambda_2|\tilde{y}_j\rangle+\Omega_2|y_j\rangle)\langle y_j|\frac{I-\sigma_3}{2}\sigma_2\\
&=\frac{4\zeta_j^2}{(\zeta_j^2-\bar\zeta_j^2)^2}\sigma_2\frac{I-\sigma_3}{2}(\Lambda|\tilde{y}_j\rangle+\Omega|y_j\rangle)\langle y_j|\frac{I-\sigma_3}{2}\sigma_2\\
&=\frac{4\zeta_j^2}{(\zeta_j^2-\bar\zeta_j^2)^2}\sigma_2\frac{I-\sigma_3}{2}|z_j\rangle\langle y_j|\frac{I-\sigma_3}{2}\sigma_2,\\
\end{aligned}$$
in terms of (\ref{d9}) and (\ref{d33}). The second case of (\ref{d32}) can be proved similarly.  \quad$\blacksquare$

\begin{proposition}\label{p2}
$\tilde{B}_j(x,t)$ defined in (\ref{d7}) can also be represented in terms of $D^{-1}(\zeta)$ defined by (\ref{d3}) at the point $\zeta_j$
\begin{equation}\label{d34}
\sigma_2\tilde{B}_j^T(x,t)\sigma_2=\frac{(\zeta_j^2-\bar\zeta_j^2)^2}{4\zeta_j^2}D^{-1}(\zeta_j).
\end{equation}
\end{proposition}
Proof: For the sake of convenience, we denote the diagonal part of a matrix $A$ by $[A]^{(d)}$ and the off-diagonal part by $[A]^{(o)}$,
then $A=[A]^{(d)}+[A]^{(o)}$.
According to Lemma \ref{l3} we find that (\ref{d31}) implies that
$$[D_j^{-1}(\zeta_j)]^{(o)}=\frac{4\zeta_j^2}{(\zeta_j^2-\bar\zeta_j^2)^2}\overline{\sigma_2[\tilde{B}_j^\dagger]^{(o)}\sigma_2},$$
and (\ref{d32}) gives
$$[D_j^{-1}(\zeta_j)]^{(d)}=\frac{4\zeta_j^2}{(\zeta_j^2-\bar\zeta_j^2)^2}\overline{\sigma_2[\tilde{B}_j^\dagger]^{(d)}\sigma_2}.\qquad \blacksquare$$


\end{document}